\documentclass{article}

\begin{document}

\title{A van der Waals free energy in electrolytes revisited}

\author{B. Jancovici\thanks{e-mail: Bernard.Jancovici@th.u-psud.fr}}

\date{}

\maketitle

\noindent Laboratoire de Physique Th{\'e}orique, Universit{\'e} de
Paris-Sud, B{\^a}timent 210, 91405 Orsay cedex, France (Unit{\'e} Mixte de
Recherche 8627-CNRS)

\begin{abstract}
A system of three electrolytes separated by two parallel planes is considered.
Each region is described by a dielectric constant and a Coulomb fluid in the
Debye-H\"uckel regime. In their book Dispersion Forces, Mahanty and Ninham
have given the van der Waals free energy of this system. We rederive this free
energy by a different method, using linear response theory and the
electrostatic Maxwell stress tensor for obtaining the dispersion force. 
\end{abstract}


\vskip 0.7 truecm
\noindent LPT Orsay 05-41

\newpage

\section{Introduction}

The modest aim of the present paper is to give another derivation for the van
der Waals free energy (generating a dispersion force) in a macroscopic
arrangement of electrolytes. The considered arrangement is as follows: Space
is divided into three regions $i=1,2,3$ by two parallel planes distant by $l$
from each other; with the Cartesian coordinate system chosen such that the $x$
axis is normal to the planes, we define region $i=1$  by $x<-(l/2)$, region
$i=2$ by $-(l/2)<x<(l/2)$, and region $i=3$ by $x>(l/2)$. Each region may be
filled by an electrolyte. The solvent is macroscopically described by a
continuous dielectric medium with a given dielectric constant $\epsilon_i$ and
a dissolved salt microscopically modelled by a Coulomb fluid in the
Debye-H{\"u}ckel regime with a given Debye wavenumber $\kappa_i$. In their
book~\cite{Ninham}, Mahanty and Ninham write, for this arrangement, a free
energy, per unit area of one plane, of the form 
\begin{equation}
F(l)=\frac{k_{\rm B}T}{2}\int_0^{\infty}\frac{2\pi k {\rm d}k}{(2\pi)^2}
\ln[1-\Delta_{12}\Delta_{13}\exp(-2s_2l)],    \label{freeenergy}
\end{equation}
where $k_{\rm B}$ is Boltzmann's constant, $T$ the temperature, 
$s_i^2=\kappa_i^2+k^2$, and $\Delta_{ij}=(\epsilon_is_i-\epsilon_js_j)/
(\epsilon_is_i+\epsilon_js_j)$. If $\kappa_2=0$, $F(l)$ is long-ranged;
if $\kappa_2\neq 0$, $F(l)$ is ``screened'' into a short-range expression. 
The result (\ref{freeenergy}) has many applications in electrochemistry,
biology, etc.

In ref.~\cite{Ninham} (see also~\cite{Ninham1}), the authors start with a
quantum system. They work out a quantum elaborate formalism, in which the
$l$-dependent part of the free energy is written as a sum on Matsubara
frequencies $\xi_n=2\pi nk_{\rm B} T/\hbar$, where $\hbar$ is Planck's
constant (divided by $2\pi$). Actually, eq.(\ref{freeenergy}) is the $\xi_n=0$
term of that sum. In the high-temperature regime, the quantum free energy
should go to the classical one, and only the $\xi_n=0$ term (\ref{freeenergy})
should survive. Therefore, a derivation of (\ref{freeenergy}), using from the
start equilibrium classical (i.e. non-quantum) statistical mechanics, should
be possible. Such a derivation has been given by Netz in a field-theoretical
formalism~\cite{Netz}. Here, we  present another purely classical derivation,
based on the use of linear response theory and the Maxwell stress tensor,
generalizing a method already used in a special case~\cite{Janco}; in this
way, we obtain the dispersion force corresponding to (\ref{freeenergy}). 

In classical statistical mechanics, there is no coupling between charges and
electromagnetic radiation (Bohr-van Leeuwen 
theorem)~\cite{Alastuey,Buenzli,Janco1}. Therefore,  we have to consider only
charges interacting by the static Coulomb law.

We model each electrolyte as a system of $M_i$ species $\alpha_i=1,2,...,M_i$
of particles, each of them carrying a charge $q_{\alpha_i}$. The fugacity of
species $\alpha_i$ is given as $z_{\alpha_i}$. If each electrolyte occupied the
whole space, the number density of species $\alpha_i$ would be $n_{\alpha_i}$.
The Debye wavenumber $\kappa_i$ in region $i$ is defined by
\begin{equation}
\kappa_i^2=\frac{4\pi\beta\sum_{\alpha_i}n_{\alpha_i}q_{\alpha_i}^2}
{\epsilon_i},     \label{kappa}
\end{equation}
where $\beta=1/(k_{\rm B} T)$.

The paper is organized as follows. In Section 2, we compute the electric
potential created by an unit charge. In Section 3, we use the linear
response theory for computing the Maxwell electrostatic stress tensor. In
Section 4, we compute the kinetic part of the stress tensor. Finally, in
Section 5, we obtain the total stress, and, by integration, recover the
Mahanty-Ninham free energy (\ref{freeenergy}). 

\section{A Green function}

In the following, we shall need the electric potential at ${\bf r}_1$
knowing that there is a unit charge at ${\bf r}_2$. In the Debye-H\"uckel
regime (which describes the small-$\beta$ case) this potential 
$G({\bf r}_1,{\bf r}_2)$ is the Green function obeying 
\begin{equation}
(\Delta_1-\kappa_i^2)G({\bf r}_1,{\bf r}_2)
=-(4\pi/\epsilon_j)\delta({\bf r}_1-{\bf r}_2),   \label{Green}   
\end{equation}
where the index $i$ denotes the region where is ${\bf r}_1$ and $j$ denotes
the region where is ${\bf r}_2$. At the lowest relevant order in $\beta$, the
densities can be replaced by the bulk ones, this is why the bulk $\kappa_i$
are used in (\ref{Green}). This equation is supplemented by the continuity
conditions at the interfaces of $G$ and $\epsilon(\partial G/\partial x_1)$,
and by the condition that $G$ should vanish when $|x_1|\rightarrow\infty$
(this last condition means that we assume that there is no bulk potential
difference between regions 1 and 3, and these potentials have been chosen as
zero without loss of generality). 

Eq.(\ref{Green}) can be replaced by a simpler one for the transverse Fourier
transform of $G$. Let ${\bf r}^{\perp}$ be the two-dimensional vector which
describes the components of ${\bf r}$ normal to the $x$ axis. Since $G$
depends on ${\bf r}^{\perp}_1$ and ${\bf r}^{\perp}_2$ only through their
difference, we can define their transverse Fourier transform by
\begin{equation}
{\tilde G}(x_1,x_2,k)=\int{\rm d}({\bf r}^{\perp}_1-{\bf r}^{\perp}_2)
\exp[-{\rm i}{\bf k}\cdot({\bf r}^{\perp}_1-{\bf r}^{\perp}_2)]
G({\bf r}_1,{\bf r}_2),   \label{Fourier}
\end{equation}
where $k=|{\bf k}|$ is the modulus of a two-dimensional wavenumber normal to
the $x$ axis. Eq.(\ref{Green}) becomes a simpler differential equation. We
shall only need the case when $x_2$ is in region 2. Then, this equation is
\begin{equation}  
(\frac{\partial^2}{\partial x_1^2}-k^2-\kappa_i^2){\tilde G}(x_1,x_2,k)
=-(4\pi/\epsilon_2)\delta(x_1-x_2).    \label{tildeGreen}
\end{equation}
The solution of this equation which vanishes when $|x_1|\rightarrow\infty$
is of the form
\begin{eqnarray} 
& &{\tilde G}=\frac{2\pi}{\epsilon_2s_2}\exp(-s_2|x_1-x_2|)+A(x_2)\exp(-s_2x_1)
\nonumber \\
& &+B(x_2)\exp(s_2x_1)\;\; {\rm when}\;-(l/2)<x_1<(l/2),   \nonumber  \\
& &{\tilde G}=C(x_2)\exp(s_1x_1)\;\; {\rm when}\;x_1<-(l/2),
\label{tildeGreen1}\\
& &{\tilde G}=D(x_2)\exp(-s_3x_1)\;\; {\rm when}\;x_1>(l/2), \nonumber
\end{eqnarray} 
where $s_i^2=\kappa_i^2+k^2$. The coefficients $A,B,C,D$ are to be determined
by the continuity conditions at the interfaces. One finds, when both $x_1$ and
$x_2$ are in region 2, ${\tilde G}={\tilde G}_{\rm bulk}+{\tilde G}_1$, where 
${\tilde G}_{\rm bulk}(x_1,x_2,k)=(2\pi/\epsilon_2s_2)\exp(-s_2|x_1-x_2|)$ is
the Fourier transform of the bulk Debye screened potential 
$G_{\rm bulk}({\bf r}_1,{\bf r}_2)=(\exp(-\kappa_2|{\bf r}_1-{\bf r}_2|)/
(\epsilon_2|{\bf r}_1-{\bf r}_2|)$ and ${\tilde G}_1$, the Fourier transform of
$G_1$, is
\begin{eqnarray}
& &{\tilde G}_1(x_1,x_2,k)=\frac{2\pi}{\epsilon_2s_2}
\frac{2\Delta_{12}\Delta_{32}\exp(-2s_2l)\cosh[s_2(x_1-x_2)]}
{1-\Delta_{12}\Delta_{32}\exp(-2s_2l)} \nonumber  \\
& &-\frac{2\pi}{\epsilon_2s_2}\frac{\exp(-s_2l)\{\Delta_{12}\exp[-s_2(x_1+x_2)]
+\Delta_{32}\exp[s_2(x_1+x_2)]\}}{1-\Delta_{12}\Delta_{32}\exp(-2s_2l)}. 
\label{tildeGreen2}
\end{eqnarray}

\section{Maxwell stress tensor}

In statistical mechanics $qG({\bf r}_1,{\bf r}_2)$  means the statistical
average of the total electric potential at ${\bf r}_1$ when an  additional
charge $q$ is placed at ${\bf r}_2$. Therefore, in terms of the microscopic
electric potential ${\hat\phi}({\bf r})$ created by the system (dielectrics
plus electrolytes, NOT $q$),  
\begin{equation}
<{\hat\phi}({\bf r}_1)>_q
=qG({\bf r}_1,{\bf r}_2)-\frac{q}{|{\bf r}_1-{\bf r}_2|}, \label{response}
\end{equation}
where $<\ldots>_q$ is a statistical average taken when $q$ is present. The
presence of $q$ adds to the Hamiltonian the term ${\hat H}'=
q{\hat\phi}({\bf r}_2)$. In the classical regime, assumed to be valid for both
the charged particles and the dielectrics, linear response theory says that,
when $q$ is infinitesimal 
\begin{equation}
<{\hat\phi}({\bf r}_1)>_q=-\beta<{\hat H}'{\hat\phi}({\bf r}_1)>,
\label{linear}
\end{equation}
where $<\ldots>$ is a statistical average taken without ${\hat H}'$ in the 
statistical weight. In writing (\ref{linear}), we have taken into account that
$<{\hat\phi}({\bf r}_1)>=0$ at the lowest relevant order in $\beta$; indeed,
if the rhs of eq.(\ref{Green}) is replaced by zero, its solution vanishing at
infinity is identically zero. We shall use (\ref{response}) and (\ref{linear})
for computing the $l$-dependent part of the Maxwell stress tensor. We define a
$l$-dependent part by the condition that it vanishes when
$l\rightarrow\infty$. For obtaining that $l$-dependent part, it can be checked
that it is sufficient to replace the rhs of (\ref{response}) by its part
$qG_1({\bf r}_1,{\bf r}_2)$. From now on, all quantities refer to their
$l$-dependent part. (\ref{response}) and (\ref{linear}) give   
\begin{equation}
\beta <{\hat\phi}({\bf r}_1){\hat\phi}({\bf r}_2)>=-G_1({\bf r}_1,{\bf r}_2).
\label{correlation}
\end{equation}

The $xx$ component of the Maxwell stress tensor\footnote{Since here the
electric potential is the microscopic one due to the charged particles of the
system, not the macroscopic one due to some external charges, the simple 
Minkowski expression of the stress tensor is appropriate} in region 2 
is~\cite{Jackson} 
\begin{equation}
T_{xx}(x_1)=\frac{\epsilon_2}{4\pi}\left(\frac{\partial^2}
{\partial x_1\partial x_2}-\frac{1}{2}\nabla_1\cdot\nabla_2\right)
<{\hat\phi}({\bf r}_1){\hat\phi}({\bf r}_2)>_{{\bf r}_2={\bf r}_1},
\label{Maxwell}
\end{equation}
or, in terms of the Fourier transform of (\ref{correlation}),
\begin{eqnarray}
& &\beta T_{xx}(x)=-\frac{\epsilon_2}{8\pi}\int\frac{{\rm d}{\bf k}}{(2\pi)^2}
\left(\frac{\partial^2}{\partial x_1\partial x_2}-k^2\right)
{\tilde G}_1(x_1,x_2,k)_{x_1=x_2=x} \nonumber \\  
& &=\frac{1}{4}\int\frac{{\rm d}{\bf k}}{(2\pi)^2s_2}
\left[\frac{(\kappa_2^2+2k^2)2\Delta_{12}\Delta_{32}\exp(-2s_2l)}
{1-\Delta_{12}\Delta_{32}\exp(-2s_2l)}\right. \nonumber \\
& &+\left.\frac{\kappa_2^2\exp(-s_2l)[\Delta_{12}\exp(-2s_2x)
+\Delta_{32}\exp(2s_2x)]}
{1-\Delta_{12}\Delta_{32}\exp(-2s_2l)}\right]. \label{Maxwell1}
\end{eqnarray} 

\section{Kinetic part of the stress tensor}

The $xx$ component of the stress tensor has, in addition to its Maxwell part, 
a kinetic part, which is, in region 2, $-k_BTn_2(x)$ where $n_2(x)$ is the
total number density of the electrolyte at $x$. It happens that the departure
of $n_2(x)$ from its bulk value $n_2=\sum_{\alpha_2}n_{\alpha_2}$ gives to the
$l$-dependent part of the stress tensor a contribution of the same order than
the Maxwell part, although the Maxwell part can be and has been computed, at
the lowest order in $\beta$, by using the bulk densities in (\ref{Green}). 

For computing $n_2(x)$, a natural idea is to use the first equation of the YBG
hierarchy which gives ${\rm d}n_2(x)/{\rm d}x$; however, it is not easy to find
the integration constant for getting $n_2(x)$. We resort to a slightly more
complicated method relying on the diagrammatic renormalized Mayer expansion
reviewed in ref.~\cite{Samaj}. The Debye-H\"uckel regime is described by
writing for the excess free energy $F_{\rm exc}$  only the diagrams 
\begin{equation}
-\beta F_{\rm exc}= \bullet---\bullet + {\rm ring}\;{\rm diagrams}\;,
\label{excfreeenergy}
\end{equation}
where a black circle stands for the density of some species of particles at
some position (to be summed and integrated on), a bond stands for $-\beta$
times the Coulomb interaction between the particles at its ends, and the ring
diagrams are rings made of $n$ ($n\geq 2$) black circles and $n$ bonds. Taking
the functional derivative of (\ref{excfreeenergy}) with respect to 
$n_{\alpha_2}({\bf r})$ gives
\begin{eqnarray} 
& &\ln\left[\frac{n_{\alpha_2}(x)}{z_{\alpha_2}}\right]=\circ---\bullet\:+
\nonumber \\
& &\frac{1}{2}(\circ---\bullet---\circ\:+\:
\circ---\bullet---\bullet---\circ\:+\:\ldots)\:, \label{density}
\end{eqnarray}
where all white circles refer to the same fixed particle of species $\alpha_2$
at some point ${\bf r}$ in region 2 (actually the densities depend only on the
$x$ component of ${\bf r}$). In the rhs of (\ref{density}), the first diagram
is $-\beta q_{\alpha_2}<{\hat\phi}(x)>$. The other diagrams are the sum of
chain diagrams minus the first one $\circ---\circ$; $-\beta$ times the
screened potential $q_{\alpha_2}^2G$ is known to be given by this sum of chain
diagrams (taking into account the position-dependence of the densities in the
calculation of $G$ would give corrections of higher order in
$\beta$). Therefore, 
\begin{equation}
\ln\left[\frac{n_{\alpha_2}(x)}{z_{\alpha_2}}\right]  
=-\beta q_{\alpha_2}<{\hat\phi}(x)>-\frac{\beta q_{\alpha_2}^2}{2}
\lim_{{\bf r}'\rightarrow{\bf r}}\left[G({\bf r},{\bf r}')
-\frac{1}{\epsilon_2|{\bf r}-{\bf r}'|}\right]. \label{density1}
\end{equation}
For the bulk values, a similar equation is obtained, by replacing in
(\ref{density1}) $n_{\alpha_2}(x)$ by $n_{\alpha_2}$, $<{\hat\phi}(x)>$ by
$<{\hat\phi}_{\rm bulk}>$, and $G$ by $G_{\rm bulk}$. Combining these two
equations, we obtain  
\begin{equation}
n_{\alpha_2}(x)=n_{\alpha_2}
\exp\left[-\beta q_{\alpha_2}[<{\hat\phi}(x)>-<{\hat\phi}_{\rm bulk}>]
-\frac{\beta q_{\alpha_2}^2}{2}G_1({\bf r},{\bf r})\right] \label{density2}
\end{equation}
Linearizing the exponential in (\ref{density1}), summing on $\alpha_2$, taking
into account the bulk neutrality condition
$\sum_{\alpha_2}q_{\alpha_2}n_{\alpha_2}=0$ which supresses the term
containing $<{\hat\phi}(x)>-<{\hat\phi}_{\rm bulk}>$, and using
(\ref{tildeGreen2}), we obtain 
\begin{eqnarray}
& &n_2(x)-n_2=-\frac{\kappa_2^2}{4}\int\frac{{\rm d}{\bf k}}{(2\pi)^2s_2}
\left[\frac{2\Delta_{12}\Delta_{32}\exp(-2s_2l)}
{1-\Delta_{12}\Delta_{32}\exp(-2s_2l)}\right. \nonumber \\
& &-\left.\frac{\exp(-s_2l)[\Delta_{12}\exp(-2s_2x)
+\Delta_{32}\exp(2s_2l)]}{1-\Delta_{12}\Delta_{32}\exp(-2s_2l)}\right]. 
\label{density3}
\end{eqnarray}
Since $n_2(x)-n_2\rightarrow 0$ as $l\rightarrow\infty$, the $l$-dependent
part of $n_2(x)$ is $n_2(x)-n_2$. 

\section{Final result}

Finally, by adding the Maxwell and kinetic contributions in region 2, one
finds, for the $l$-dependent part of the force  per unit area acting on the
region on the right of $x$, $f=-T_{xx}(x)+k_{\rm B}T[n_2(x)-n_2]$. Thus,  
\begin{equation}
f=-\frac{k_{\rm B}T}{2}\int_0^\infty\frac{{\rm d}{\bf k}}{(2\pi)^2}
\frac{2s_2\Delta_{12}\Delta_{32}\exp(-2s_2l)}
{1-\Delta_{12}\Delta_{32}\exp(-2s_2l)}.  \label{force}
\end{equation}
This force is, as it should be, independent of $x$. This general expression
has been rederived in special cases, sometimes by different methods, see for
instance ref.~\cite{Ninham2,Janco}. Since $f=-\partial F(l)/\partial l$,
integration of $f$ with respect to $l$, with the integration constant
determined by the condition that $F$ vanishes as $l\rightarrow\infty$, gives
(\ref{freeenergy}). 

In the formalism used by Netz~\cite{Netz}, (\ref{freeenergy}) has been
obtained in a ``one-loop'' approximation; a ``two-loop'' correction has been
computed in a special case. An alternative approach for obtaining corrections 
would be to extend the calculation in the present paper by taking into account
higher-order terms in the renormalized Mayer expansion. We do not know which
approach is the simplest one.

\end{document}